\documentclass[prb,10pt,twocolumn,amsmath,amssymb,showpacs,floatfix]{revtex4}

\usepackage{graphicx}
\usepackage{amssymb, amsmath}
\usepackage{dcolumn}

\newcommand{\dd}{~\mathrm{d}}
\newcommand{\be}{\begin{equation}}
\newcommand{\ee}{\end{equation}}
\newcommand{\bea}{\begin{eqnarray}}
\newcommand{\eea}{\end{eqnarray}}
\newcommand{\bean}{\begin{eqnarray*}}
\newcommand{\eean}{\end{eqnarray*}}

\newcommand{\vc}[1]{{\bf #1}}

\def\k3picene{K$_3$-picene}

\begin{document}
\title{Local and non-local electron-phonon couplings in \k3picene and the
effect of metallic screening.}

\author{Michele Casula, Matteo Calandra, and Francesco Mauri}
\affiliation{CNRS and Institut de Min\'eralogie et 
de Physique des Milieux condens\'es,
Universit\'e Pierre et Marie Curie,
case 115, 4 place Jussieu, 75252, Paris cedex 05, France}

\date{\today}
\begin{abstract} 
We analyze the properties of electron-phonon couplings in
\k3picene by exploiting a molecular orbital representation derived  in
the maximally localized Wannier function formalism. 
This allows us to go beyond the analysis done in Phys. Rev. Lett. 
{\bf 107}, 137006 (2011), and separate not only the intra- and
intermolecular phonon contributions but also the
local and non-local electronic states in the electron-phonon matrix elements.
Despite the
molecular nature of the crystal, we find that the
purely molecular contributions (Holstein-like couplings where the
local deformation potential is coupled to intramolecular phonons) account for only $20\%$ of the total
electron-phonon interaction $\lambda$. 
In particular, the Holstein-like contributions to $\lambda$ in
\k3picene are four times smaller than those computed for an isolated
neutral molecule,
as they are strongly screened by the metallic bands of the doped crystal.
Our findings invalidate the use of molecular electron-phonon
calculations to estimate the total electron-phonon coupling in
metallic picene, and possibly in other doped metallic molecular crystals.
The major contribution ($80\%$) to $\lambda$ in \k3picene comes from non-local
couplings due to phonon modulated hoppings. 
We show that the
crystal geometry together with the molecular picene structure leads
to a strong 1D spatial anisotropy of the non-local
couplings. Finally, based on the parameters derived from our
density functional theory calculations, we propose a lattice modelization of the 
electron-phonon couplings in \k3picene which gives $90\%$ of ab-initio
$\lambda$.
\end{abstract}
\pacs{31.15.A-, 74.70.Kn, 63.20.kd}

\maketitle

\section{Introduction}
\label{intro}

Understanding transport properties of molecular crystals
based on hydrocarbon molecules is relevant not only to fundamental
condensed matter physics, but also for applications in nanoelectronics.
For instance, organic field-effect transistors are appealing as they
are flexible, lightweight and cheap. Rubrene-based field effect transistors \cite{Sundar07,Takeya07}
display tunable mobilities that can be as large as 40 cm$^2$/(V
$\cdot$ s). More recently it was shown that picene field-effect transistors\cite{OkamotoPiceneFet} based on liquid
electrolytes have p-channel characteristics\cite{Kawai2012},
although with much reduced mobilities with respect to rubrene.

Transport properties of
organic molecular crystals can also be tuned
by intercalation of alkali or alkaline earth metals. 
K intercalation leads to metallic
states in phthalocyanine materials \cite{Cracium06} and in 
several other polycyclic aromatic hydrocarbons.
In picene\cite{Mitsuhashi2010}, phenanthrene\cite{Wang2011,Wang2011b},
coronene\cite{KubozonoCoronene}, and in 1,2:8,9-dibenzopentacene\cite{Xue2011},
intercalation stabilizes a superconducting state with critical
temperatures (T$_c$) up to 33 K.
A detailed understanding of transport phenomena in such systems
is then relevant also for the realm of fundamental research.

An important source of intrinsic scattering in aromatic molecular crystals is
provided by the electron-phonon coupling. In these systems there is an interplay between
intramolecular local interactions and intermolecular non-local
interactions. Determining the mutual role of local and non-local 
interactions is hardly doable without a proper theoretical
approach. Molecular crystals can indeed behave very differently
depending on the details of the molecules composing the crystal and 
on their arrangement.
In alkali doped fullerenes\cite{GunnarssonRMP}, superconductivity 
is supposed to be mostly due to intramolecular phonons.
In this case, if the electronic states coupled to the phonons
are molecular and the metallic screening is weak, then the problem can be tackled at a 
molecular level by the calculation of electron-phonon interaction
\cite{Janssen2010,Faber2011} on an isolated ionized molecule. 

The situation is more complicated in the field of hydrocarbon
molecular crystals. In the case of \k3picene, molecular calculations
\cite{Kato2011} using the B3LYP functional give a large electron-phonon coupling, that can
almost alone explain T$_c$. However, the generalization of
this approach to other hydrocarbon molecular crystals predicts
a decrease of the critical temperature with the increase of the
molecular size, in disagreement with experimental 
data~\cite{Mitsuhashi2010,Wang2011,Wang2011b,Xue2011}.
Indeed in experiment the largest T$_c$ is for the crystal composed
by the largest molecules.
Subedi {\it et al.}~\cite{Subedi} performed a
density functional theory calculations(DFT) in which the crystal
structure of pristine picene was adopted and K-doping was treated
in a rigid doping approach. The screening of the self-consistent
potential was assumed to be that of
insulating picene. A very large
electron-phonon coupling was found mostly due to intramolecular
phonons, in agreement with Ref.~\onlinecite{Kato2011}. 

In our previous work \cite{picene_prl_casula} we performed DFT
calculations relying on less approximations then in
Ref.~\onlinecite{Subedi}. The theoretically devised crystal structure of 
\k3picene was considered \cite{picene_prl_casula,Kosugi09} 
and K-atoms were explicitly included in the
calculation. Furthermore we included 
the metallic screening of crystalline \k3picene
in the self-consistent potential.
By projecting the phonon polarizations into intramolecular
and intermolecular vibrations, we found that \k3picene has a strong electron-phonon coupling 
($\lambda=0.73$) that is partially due to the coupling to intermolecular
and intercalant phonons ($40\%$) and partially to the coupling to intramolecular
phonons ($60\%$),
in disagreement with Refs.~\onlinecite{Kato2011,Subedi}.
In our present work, we go beyond what we have done in
Ref.~\onlinecite{picene_prl_casula}. Instead of analyzing the
``locality'' of the electron-phonon coupling in terms of phonon
projections only, we study it also by means of electronic projections onto a molecular basis, which allows
one to distinguish between the on-site electronic Hamiltonian
and the hopping parts, both modulated by the coupling with phonons.
This approach leads to a stricter distinction between purely molecular and
crystal contributions, and yields a further reduction of the
purely molecular component, estimated to be
about $20\%$  of the total $\lambda$.

The three approaches illustrated above, namely molecular calculations,
rigid doping of the crystal, and explicit treatment of the dopants, rely on different approximations
that could explain the discrepancies. An important one is the treatment
of the electronic screening and its effects on the electron-phonon
interaction. In molecular calculations \cite{Kato2011} and in
Ref.~\onlinecite{Subedi}, metallic screening is 
neglected.
Analogy with alkali doped fullerenes points out, however, that this
assumption is not necessary fulfilled.
In K$_3$C$_{60}$, it has been suggested that metallic screening 
strongly affects the electron-phonon coupling
\cite{Zwicknagl92,Zwicknagl92b,Antropov}.
For example, A$_{1g}$ modes causing a shift without
splitting of the t$_{1u}$ C$_{60}$ molecular levels, are supposed
to be screened by the charge transfer from up-shifted to down-shifted levels\cite{Antropov},
i. e. by the metallic screening in the solid.
In \k3picene the situation could be similar. 
However, the relative contribution of
intramolecular, intermolecular and intercalant interactions
remains unclear, and largely unexplored.
In this work we carry out a detailed and quantitative analysis 
of the total electron-phonon coupling $\lambda$ in \k3picene,
by addressing these issues. 

The paper is organized as follows. In Sec.~\ref{definition} we provide
the general definition of local and non-local electron-phonon
couplings for a molecular crystal. In Sec.~\ref{wannier} we describe
the geometry of \k3picene, and we show the dominant hoppings of the
corresponding tight-binding Hamiltonian in the
Wannier basis. In Sec.~\ref{local} electron-phonon calculations
are carried out by discriminating between local and
non-local couplings. We find that 
the purely local contributions account for only $20\%$ of the full
$\lambda$, while the remaining part comes from
non-local sources.
Sec~\ref{mol_screening} analyzes the screening acting on
the local electron-phonon terms by a direct comparison between the
crystal and the isolated (unscreened) molecule. We show that 
the effect of the metallic screening provided by the crystal environment
to the deformation potential is sizable with a strong reduction of the  
local electron-phonon coupling with respect to the corresponding
strength found in the neutral isolated molecule.
In Sec.~\ref{dimg} we look for the most important non-local terms
contributing to the total $\lambda$ and we build a model Hamiltonian
with few non-local electron-phonon couplings added to the local
part which gives $90\%$ of the total $\lambda$. The conclusions are in
Sec.~\ref{conclusions}.

\section{Definition of local and non-local electron-phonon couplings
  in a molecular crystal}
\label{definition}

We suppose that the band structure of a molecular crystal is described
by the electronic tight-binding Hamiltonian $H_\textrm{el}$ written in
a basis set built out of molecular orbitals $|i , m \rangle =
c^\dagger_{im} | 0 \rangle$, where $i$ is the index of the molecular site
having its center of mass located at the equilibrium position ${\bf R}_i$,
and $m$ is the orbital index, with $c^\dagger_{im}$ and $c_{jn}$
satisfying canonical anticommutation relations.
For simplicity, we assume here that there is only one molecule per unit
cell, so the vectors ${\bf R}_i$ define also the Bravais lattice.  
$H_\textrm{el}$ reads then as
\begin{equation}
H_\textrm{el}=-\sum_{ij} \sum_{mn} t_{mn}({\bf R}_j-{\bf R}_i) c^\dagger_{im} c_{jn},
\label{Hel}
\end{equation}
where we omitted the spin index by implicitly assuming that the spin up and spin down components
are equivalent, namely there is no spin symmetry breaking. The hopping
matrix is defined as:
\begin{equation}
-t_{mn}({\bf R}_j-{\bf R}_i) = \langle  i,m | H | j,n \rangle,
\end{equation}
where we exploit the lattice translational invariance.

In second quantization a phonon displacement ${\bf u}_{s}$ of atom $s$ 
with mass $M_{s}$ relative to the i-th molecule is:
\begin{equation}
{\bf u}_{s}({\bf R}_i)=\frac{i}{N_{\bf q}} \sum_{{\bf q}\nu} 
\frac{1}{\sqrt{2 M_{s}\omega_{{\bf q}\nu}}} {\bf e}_{{\bf q}\nu}^{s}(b_{{\bf
     q}\nu}+b_{-{\bf q}\nu}^{\dagger} )
e^{i{\bf q}\cdot{\bf R}_i+i {\bf q}\cdot{\boldsymbol \tau}_s}
\end{equation}
where 
$N_{q}$ is the number of phonon momentum
points describing the system, 
$\omega_{\vc{q}\nu}$ is the phonon dispersion of mode  $\nu$
at a given momentum $\vc{q}$, 
$\vc{e}_{{\bf q}\nu}^{s}$ is the
3-dimensional $s$-atomic component of the phonon eigenvector ${\bf
  e}_{{\bf q}\nu}$,
and ${\boldsymbol \tau}_s$ is the position of the atom $s$ in the unit cell.  
The operators $b^\dagger_{\vc{q}\nu}$ and $b_{\vc{q}\nu}$ satisfy
canonical bosonic commutation relations.

The harmonic phonon Hamiltonian  $H_\textrm{phon}$ reads as:
\begin{equation}
H_\textrm{phon}=\sum_{\vc{q}\nu}
\omega_{\vc{q}\nu} \left(b^\dagger_{\vc{q}\nu} b_{\vc{q}\nu}+\frac{1}{2}\right),
\label{Hphon}
\end{equation}  

The complete Hamiltonian for the electron-phonon (el-phon)  problem
includes electron-phonon
coupling terms, and is written as 
\begin{equation}
H =H_\textrm{el}  + H_\textrm{phon} + H^\textrm{local}_\textrm{el-phon} +
H^\textrm{non local}_\textrm{el-phon}
\end{equation}
where the local electron-phonon
coupling is
\begin{equation}
H^\textrm{local}_\textrm{el-phon} = \frac{1}{N_\vc{q}} \sum_{\vc{q}\nu}
\sum_i \sum_{mn} g^{\vc{q}\nu}_{mn}({\bf 0}) 
e^{i\vc{q} \cdot \vc{R}_i } c^\dagger_{im} c_{in} (b^\dagger_{-\vc{q}\nu} +
b_{\vc{q}\nu}),
\label{local_coupling}
\end{equation} 
while the non-local coupling $H^\textrm{non local}_\textrm{el-phon}$ is
\begin{equation}
 \frac{1}{N_\vc{q}} \sum_{\vc{q}\nu}
\mathop{\sum_{ij}}_{i \ne j} \sum_{mn} \left ( g^{\vc{q}\nu}_{mn}({\bf
  R}_j-{\bf R}_i) e^{i\vc{q}\cdot\vc{R}_i } c^\dagger_{im} c_{jn}
b_{\vc{q}\nu} + \textrm{h.c.}\right).
\label{non_local_coupling}
\end{equation} 
The phase $e^{i\vc{q}\cdot\vc{R}_i }$ makes the total momentum conserved in the electron-phonon
scattering terms.
The electron-phonon coupling strength projected on the molecular
orbitals is defined as
\begin{equation}
g^{\vc{q}\nu}_{mn}({\bf
  R}_j-{\bf R}_i) = \sum_s
\langle i,m |  \frac{\delta v}{\delta {\bf u}_{\textbf{q}s}} | j,n
\rangle \cdot  \textbf{e}^s_{\textbf{q}\nu} 
/\sqrt{2 M_s \omega_{\textbf{q}\nu}},
\end{equation}
where $i$,$j$ are indexes of molecular sites,
${\bf u}_{\textbf{q}s}$ is the Fourier transform of the phonon
displacement ${\bf  u}_{s}(\vc{R}_i)$, and
 $\delta v/ \delta {\bf u}_{\textbf{q}s}$ is the
(screened) deformation potential.

From Eq.~\ref{local_coupling}, it is apparent that the local coupling
is a Holstein-type interaction which couples the phonons with
on-site molecular electronic terms, while 
in Eq.~\ref{non_local_coupling} the non-local
couplings modulate the hoppings $t_{mn}({\bf R}_j-{\bf R}_i)$ in $H_\textrm{el}$ via the
bosonic fields $b^\dagger_{-\vc{q}\nu}$ and $b_{\vc{q}\nu}$.
The local and non-local coupling strengths are proportional to the
deformation potential expressed in the molecular orbital basis,
centered on either the same site or two different molecules, respectively. 
By translational invariance, the strength depends only on the vector 
${\bf R}_j-{\bf R}_i$.
It is worth pointing out that in this context the definition of ``local'' and
``non-local'' couplings is purely electronic. In our previous
work\cite{picene_prl_casula}, we distinguished between the
``intermolecular'' and ``intramolecular'' contributions based on the phonon
projections. The intramolecular phonons are those having
$\vc{e}_{\vc{q}\nu}$ projected on the single molecule manyfold, while the
intermolecular phonons are those having $\vc{e}_{\vc{q}\nu}$ spanned
by the rigid molecular rototranslations together with all intercalant displacements.
Therefore, one can expand the bosonic fields $b_{\vc{q}\nu}$ into
$b^\textrm{inter}_{\vc{q}\nu}+b^\textrm{intra}_{\vc{q}\nu}$, being the
sum of intermolecular and intramolecular projections a resolution of the identity.
We thus note that cross-contributions like 
intramolecular phonons in non-local
couplings or intermolecular phonons in local couplings, are possible.
Projection of both the electronic and phononic parts guarantees the
isolation of the single molecule contribution. In this work, we are
going to use the words ``local'' and ``non-local  couplings'' to mean
the electronic molecular basis set projections as in Eqs.~\ref{local_coupling} and
\ref{non_local_coupling}, respectively, while we keep the
notation of Ref.~\onlinecite{picene_prl_casula} by using
``intramolecular'' and ``intermolecular phonons'' to refer to the
phonon projections.

\section{Geometry and band structure of K$_3$picene}
\label{wannier}

\begin{figure*}[t!h]
\includegraphics[width=2\columnwidth]{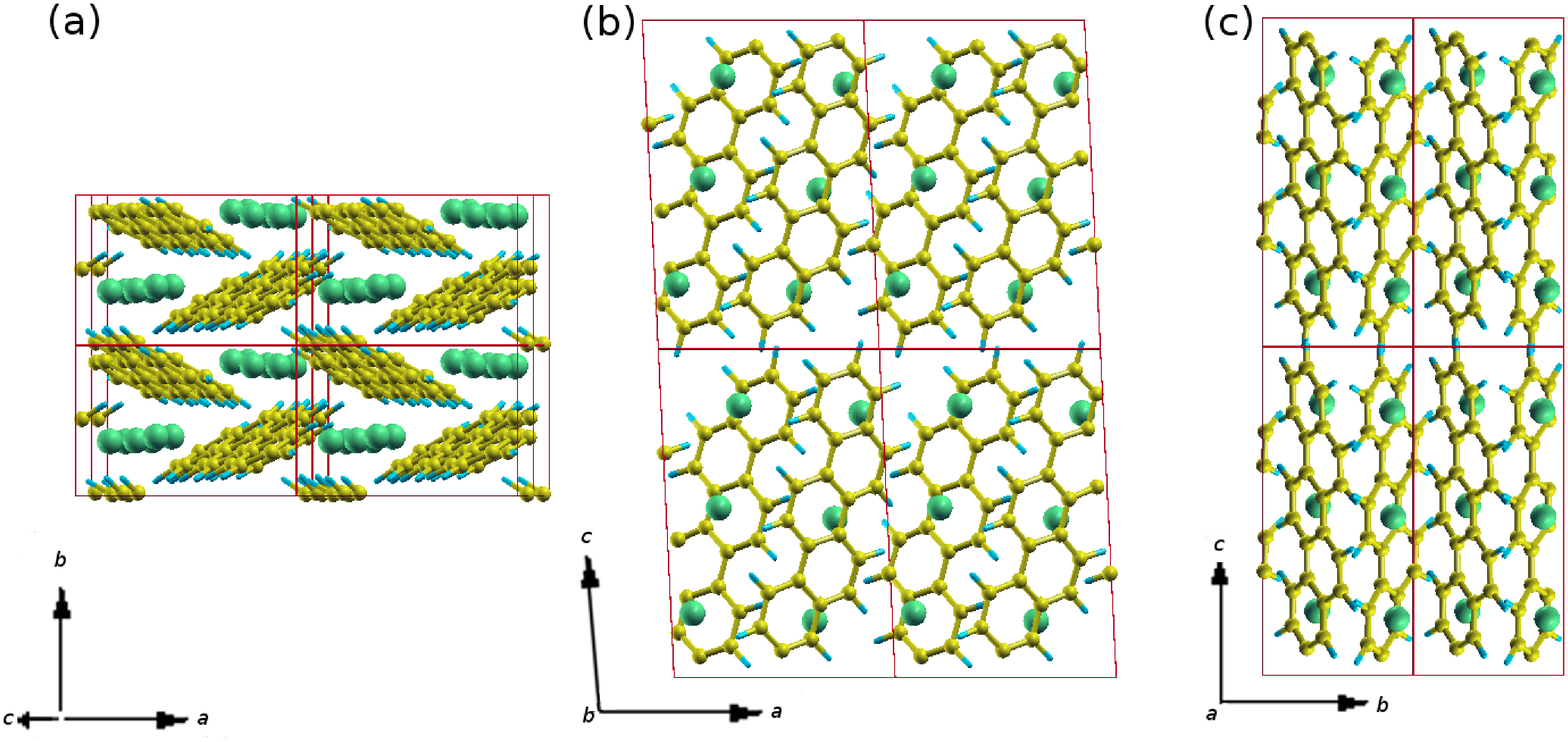}
\caption[]{(color online) Orthogonal projections of the \k3picene unit cell,
  repeated twice along each crystallographic direction. The unit cell
  sides are drawn in red. Carbon atoms are in yellow, Hydrogen
  is in blue, and Potassium in green. Panel (a): $ab$ projection with the $c$-axis
  pointing outwards; panel (b): $ac$ projection with the $b$-axis pointing
inwards; panel (c): $bc$ projection with the $a$-axis pointing
inwards\cite{xcrysden}.}
\label{layered_structure}
\end{figure*}

The molecules in the \k3picene crystal are arranged to satisfy the $P^2_1$ symmetry
group.  The unit cell contains two molecules and is monoclinic with
axes 
$a=8.707$\AA, $b=5.912$\AA, $c=12.97$\AA, 
$\alpha=90^o$,$\beta = 92,77^o$,$\gamma=90^o$. 
The unit cell parameters have been
taken from the experiment\cite{Mitsuhashi2010}, while the internal
coordinates have been
optimized after a full geometry relaxation performed in a DFT
framework within the local density approximation (LDA)
(for more details
see the supplementary materials section of
Ref. \onlinecite{picene_prl_casula}).
The final structure is drawn in Fig.~\ref{layered_structure}, where we
plot the orthogonal projections of the unit cell repeated twice in each
crystallographic direction. 
From Figs. \ref{layered_structure}(b) and \ref{layered_structure}(c),
one can clearly see the molecular stacking along the $c$-axis,
while in Fig.~\ref{layered_structure}(a) (the $ab$ projection) the
molecular herringbone arrangement of each layer is visible. The
intercalant occupies the interstitial space and tunes
the intermolecular angles by steric effect.

In order to understand the interplay between the \k3picene geometry
and its band structure, we are going to derive a tight-binding model
constructed on a Wannier function basis.
The maximally localized Wannier representation of the DFT orbitals is
useful not only to implement an interpolation scheme for computing the
band structure and the electron-phonon matrix elements, but also to
have a physical insight on the
system. For example, the formation
of the chemical bond in a solid can be visualized by means of the
Wannier representation of the molecular orbitals (MOs). In a molecular
crystal, as the picene, the Wannier representation is even more
natural, as it builds on the local nature of molecular sites,
where the MOs are strongly localized. The spatial
local representation given by the Wannier transformation helps in
modeling the electronic structure of the K-doped picene, and
understanding the mechanism which sets the
superconductivity.

By following Ref.~\onlinecite{wannier}, the maximally localized Wannier functions (MLWF) are defined as
\begin{equation}
\tilde{w}_{n\vc{R}}(\vc{r}) = 
\frac{1}{\sqrt{N_w}} \sum_\mathrm{k}
\left[ \sum_m U_{mn}^{MLWF}(\vc{k}) \psi_{m \vc{k}}(\vc{r}) \right]
e^{-i \vc{k} \cdot \vc{R}},
\label{wannier_def}
\end{equation}
where the sum $\sum_\mathrm{k}$ is over a $N_w$-point grid in the Brillouin zone
(BZ)\cite{footnote1},
$\vc{R}$ is a Bravais lattice vector, $\psi_{m
  \vc{k}}(\vc{r})$ are the Bloch eigenstates of the $m$-th band,
and $\vc{U}^{MLWF}(\vc{k})$ is a unitary matrix (for composite bands), defined
to minimize the total spread of the wave function
\begin{equation}
\Omega = \sum_n \left[ \langle \tilde{w}_{n\vc{0}} | r^2 | \tilde{w}_{n\vc{0}} \rangle - |
\langle  \tilde{w}_{n\vc{0}} | \vc{r} | \tilde{w}_{n\vc{0}} \rangle |^2 \right].
\label{spread_wannier}
\end{equation}
Note that in this case there are two molecules per unit cell, and so
the Bravais vectors $\vc{R}$ are not the centers of each molecule, at
variance with the simplest case taken into account in Sec.~\ref{definition}.
In the \k3picene,  the MLWFs have been determined for the bands
derived from the lowest unoccupied molecular orbital (LUMO), LUMO+1, and
LUMO+2 of the neutral picene molecule. Those bands form a quasi-composite group, as the LUMO is
well separated from the highest occupied molecular orbital (HOMO). The HOMO-LUMO gap in the pristine picene is 3 eV
large\cite{picene_gw}, and only the
LUMO+2 is weakly entangled with the upper bands. Therefore, a
preliminary disentanglement procedure has been
performed\cite{ivo_disentanglement}, before $\vc{U}^{MLWF}(\vc{k})$ could
be obtained. Thus, in our case $\vc{U}^{MLWF}(\vc{k})$ is a $6 \times 6$
matrix (3 bands per molecule, 2 molecules per unit cell), and a tight-binding Hamiltonian can be defined in the
rotated MLWF basis, according to the
matrix elements 
\begin{equation}
H_{nm}(\vc{R})= \langle \tilde{w}_{n\vc{0}} | H | \tilde{w}_{m
  \vc{R}} \rangle,
\label{H_wannier}
\end{equation}
where $H$ is the one-body LDA Hamiltonian.

In molecular crystals, the MLWF is not necessarily the best basis to
work with. The most ``physical'' basis is the one which
diagonalizes the local part ($H_{nm}(\vc{0})$ and $(n, m)$ running on
the same molecule)  of the Hamiltonian in
Eq.~\ref{H_wannier}. Indeed, the local part of $H$ represents the
molecule in the crystal, and its eigenvectors $w_n$ and
eigenvalues $\epsilon^\mathrm{mol}_n$ are respectively the MOs and
molecular levels in the crystal environment. From here on, 
we define $w_{n\vc{R}}(\vc{r})$ to be the ``molecular'' MLWFs, where 
$\vc{U}(\vc{k})=\vc{U}^\mathrm{MLWF}(\vc{k}) \times
\vc{U}_\mathrm{mol}$, 
with $\vc{U}_\mathrm{mol}$
being the unitary transformation which diagonalizes the local problem
in the MLWF basis. The molecular MLWFs $|w_{n\vc{R}}\rangle$ obtained
in the rigorous Wannier function formalism play the role of the
molecular orbitals $|i , m \rangle$ generically introduced in Sec.~\ref{definition}.

\begin{figure*}[t!h]
\includegraphics[width=2\columnwidth]{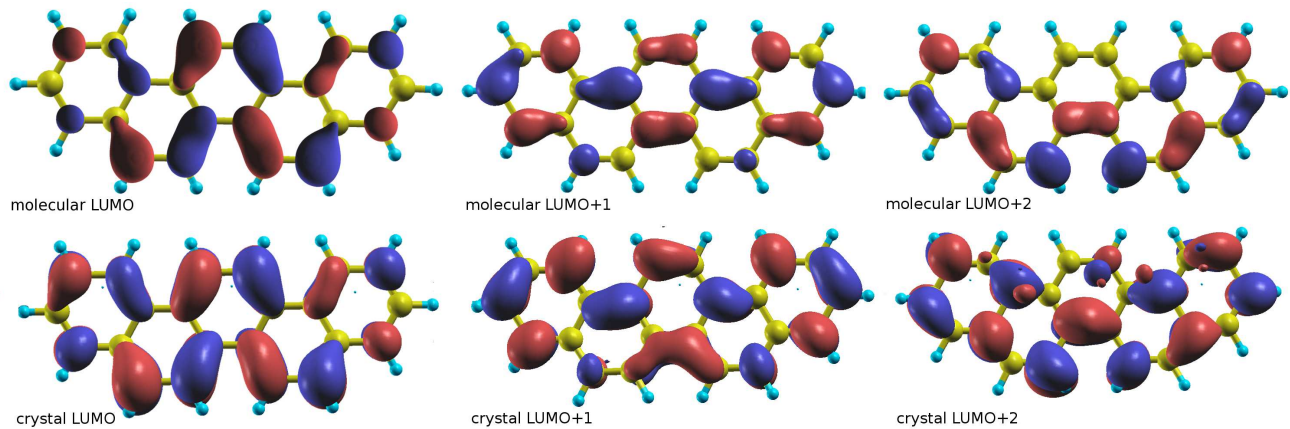}
\caption[]{(color online) Surface plot of the orbitals of an isolated
  neutral picene molecule
(upper row) and the molecular MLWF functions ($w_{n\vc{0}}(\vc{r})$)
in the \k3picene crystal (lower row).
The surface is defined by the
set of points which satisfies the condition $|\Psi_n(\vc{r})|=0.05$,
where $\Psi_n(\vc{r})= \textrm{Re}[\phi_n(\vc{r})] \textrm{Sign}[\phi_n(\vc{r})]$, with
$\phi_n(\vc{r})$ the MO whose phase has been fixed and normalized
such that  $\phi_n(\vc{r}_\mathrm{max})=1$, being
$\vc{r}_\mathrm{max}$ the location of the maximum of its modulus\cite{xcrysden}.}
\label{wannier_orbitals}
\end{figure*}

In order to see how the crystal environment affects the local
MOs, in Fig.~\ref{wannier_orbitals} we plotted 
the local molecular MLWF functions
$w_{n\vc{0}}(\vc{r})$ of \k3picene together with
the MOs of the isolated neutral molecule ($w^{MOL}_m(\vc{r})$).
One can see that $w_{n\vc{0}}(\vc{r})$ in the doped crystal are a good
representation of the orbitals in the isolated neutral picene.
Indeed, the LUMO and LUMO+1 are in close agreement. The LUMO+2
differs only slightly, as in the crystal it is more ``delocalized'',
something expected as in the molecular calculations its energy level
is close to the free particle continuum, and therefore it is more
affected by the environment.
The overall agreement allows one to make a one-to-one correspondence between the molecular
properties and the crystal local on-site properties expressed in the molecular MLWF.

\begin{figure*}[t!h]
\includegraphics[width=2\columnwidth]{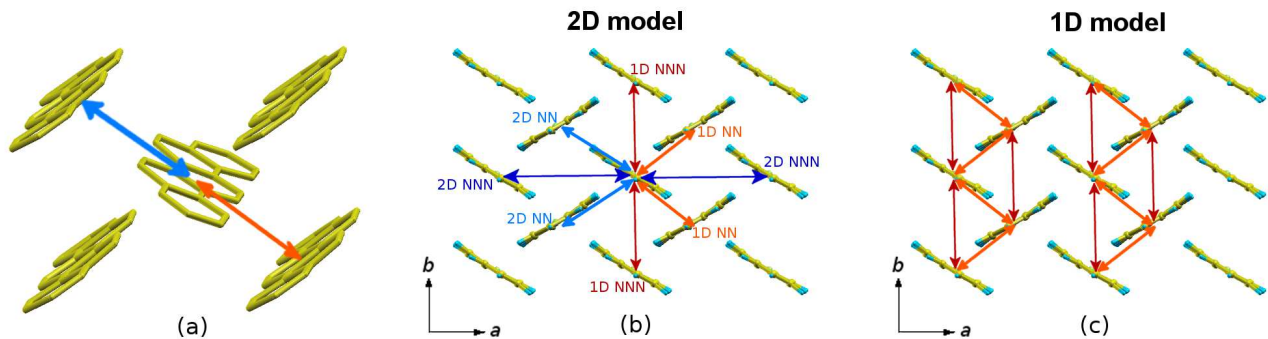}
\caption[]{(color online) Intermolecular hoppings in the herringbone
  layer. Panel (a): The two possible nearest-neighbors (NN) intermolecular hoppings are represented, with the
  strongest in orange (1D NN) mediated by a 3-ring molecular side, while the
  weakest in blue (2D NN) is bridged by a 2-ring side. Panel (b): All possible NN
and NNN hoppings in the plane are drawn. The 1D NNN are in red, while the
2D NNN are in dark blue. Panel (c): two ladder chains formed by
selecting only the 1D NN and 1D NNN hoppings, the strongest ones among the
bidimensional hoppings.}
\label{graph_structure}
\end{figure*}

We now analyze the hopping terms in Eq.~\ref{H_wannier}. They show
a clear hierarchy in magnitude depending on their spatial direction.
The largest are the
nearest neighbors (NN) hoppings which connect the molecules within the
herringbone layer. In the herringbone structure, each molecule is
linked to its four nearest neighbors in two different ways, by 
the proximity of either a two-ring molecular side, or a three-ring
side (see Fig.~\ref{graph_structure}(a)).
We found that there is a large asymmetry between the NN hoppings
connecting two molecules via a three-ring molecular side, dubbed ``1D
NN'' in the text, and the ones whose connection is bridged by a
two-ring side, dubbed ``2D NN''. The 1D NN terms, sized up to 0.09
eV, are almost twice larger than the 2D NN hoppings, which reach 0.05 eV at most. This is due to the
internal degrees of freedom of the single crystal site, as the picene
has an aromatic 5-ring structure. If the molecule were symmetric, the 1D and 2D NN hopping terms would
be equal. The consequences of this internal asymmetry will be
studied later in both the band structure and the electron-phonon couplings.

Not only the NN but also the next-nearest neighbors (NNN) hoppings are
not symmetric. Indeed, the NNN terms pointing along the $b$
crystallographic axis (named ``1D NNN'' in the text) are more than
twice larger than the ones pointing along the $a$ crystallographic
axis (dubbed ``2D NNN''), which do not go beyond 0.02 eV. This can be
easily explained by noting
that the $b$ axis is shorter than the $a$ axis, thus in the $b$
direction the molecules are more closely packed, with an increase of
the transfer integrals and so of the hoppings. See
Fig.~\ref{graph_structure}(b) for the graphical representation of all
the NN and NNN hoppings in the herringbone plane.

It turns out that the 1D NNN and the 2D NN terms are of the
same magnitude ($\approx$ 0.05 eV). The combination of 1D NN
and 1D NNN hoppings only, creates ladder chains spanning the
$b$ axis (see Fig.~\ref{graph_structure}(c)), while in a
four hopping model (with the addition of the 2D NN and the 2D NNN
terms), their combination spans the full 2D space. The 2-ring versus
3-ring asymmetry clearly favors a ``nematic'' one dimensional
electronic structure with respect to the full bidimensional
layer. Therefore we define a ``1D model''  comprising of the 0D
(on-site), 1D NN and 1D NNN terms, 
and a ``2D model'' which includes all
terms of the ``1D model'' plus the 2D NN and 2D NNN hoppings.

\begin{figure}[t!h]
\includegraphics[width=\columnwidth]{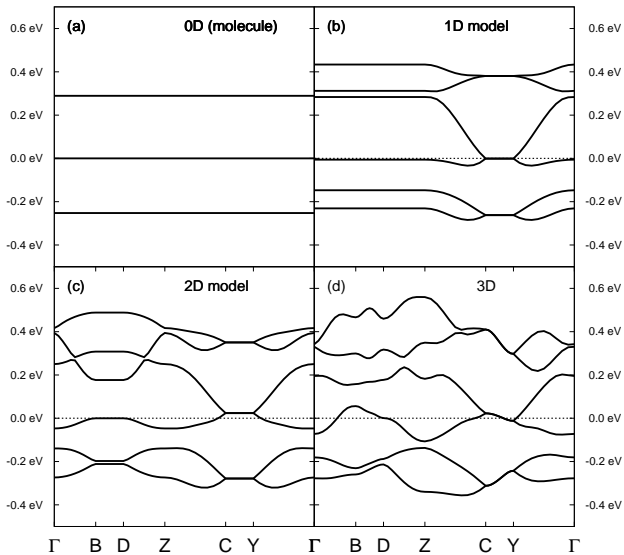}
\caption[]{Band structure of a tight-binding Hamiltonian for the
  \k3picene derived in 
  a MLWF basis including LUMO, LUMO+1 and LUMO+2 states, 
plotted along the $\Gamma$BDZCY$\Gamma$ k-space path.
 In the reciprocal crystal units, the special points are 
 $\Gamma=(0,0,0)$, B$=(\frac{1}{2},0,0)$, D$=(\frac{1}{2},0,\frac{1}{2})$,
Z$=(0,0,\frac{1}{2})$, C$=(0,\frac{1}{2},\frac{1}{2})$,
Y$=(0,\frac{1}{2},0)$.
In the left-upper panel only the ``on-site'' hoppings have been
retained, while in the right-lower panel the band structure has been
obtained with the full tight-binding model. The upper-right (the
lower-left) panel is the result of a tight-binding model with only 1D (1D+2D)
nearest neighbor and next-nearest neighbor hoppings. The zero of the
energy axis is the Fermi level.}
\label{band_structure}
\end{figure}

To understand the impact of this hierarchy on the band
structure we take the hoppings
of the tight-binding Hamiltonian written in the MLWF basis, and
we are going to selectively switch them on and off.
The full band structure is plotted in Fig.~\ref{band_structure}(d) for the
LUMO, LUMO+1, and LUMO+2 states, which yield 2 bands each.
By keeping only the local on-site terms, we obtain the
molecular levels $\epsilon^\mathrm{mol}_n$ in the crystal, which of
course are dispersionless (Fig.~\ref{band_structure}(a)). By switching
on the 1D NN and the 1D NNN hoppings along
the molecular ``wire'', one gets the band structure of the 1D model in
Fig. ~\ref{band_structure}(b). 
The dispersion develops only along the
b-axis, but it gives the main
contribution to the full 3D bandwidth, while the double degeneracy along the
CY path is due to the $P^2_1$ symmetry.
The full band structure can be roughly modeled by the 1D model, except
that the Fermi surface is poorly reproduced.
For the 2D model (see Fig.~\ref{layered_structure}(b)), a band structure closer to the 3D
one is obtained, with the LUMO bands almost perfectly reproduced, and
the flatness of the BD and CY paths due to the decoupling in the layer stacking.

\begin{figure}[t!h]
\includegraphics[width=\columnwidth]{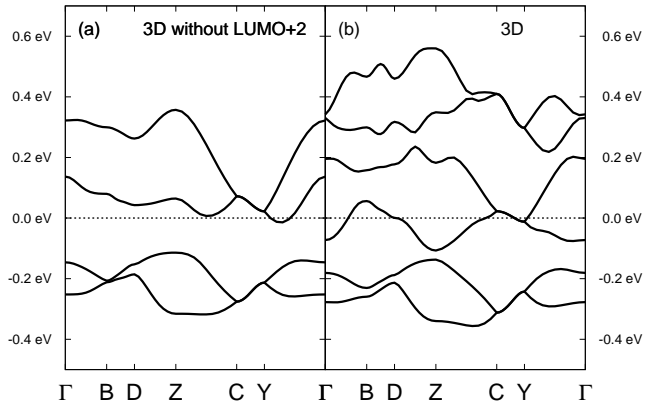}
\caption[]{The full band structure (right) and the one obtained by
  taking off the LUMO+2 states (left) from the
  tight-binding Hamiltonian obtained in the molecular MLWF basis for
  the \k3picene.
The definition of the high-symmetry points in the k-path is reported
in the caption of Fig.~\ref{band_structure}.
The zero of the energy axis is the Fermi level.
}
\label{band_structure_nolumo2}
\end{figure}

In the spirit of downfolding the full electronic structure to a
low-energy lattice model, one interesting question is whether
a two-orbital model is enough to reproduce the low-energy physics.
To this aim, we suppressed the LUMO+2 orbital from the tight-binding
model. The result is shown in Fig.~\ref{band_structure_nolumo2}. As
one can see, the LUMO+1 bands are strongly deformed, and the Fermi
surface is strongly modified. To have a correct description of the
low-energy physics of the crystal, one needs also to include the LUMO+2 molecular
orbital. Therefore, a correct modelization of the system comprises 3
orbitals, up to the LUMO+2, being the hybridization between the LUMO+1
and LUMO+2 very strong.

\subsection{Technical details for the band structure calculations}
\label{tech_dft}

The LDA-DFT calculations have been performed with the {\textsc quantum-espresso}\cite{qe}  code. K, C, and H atoms are 
described by ultrasoft pseudopotentials. The plane-wave (PW) cutoff is 60 Ry for the
wave-function, and 600 Ry for the charge.  A $4 \times 4 \times 4$
electron-momentum grid and a Methfessel-Paxton smearing  of 0.015 Ry are
used in the electronic integration. 

Wannierization has been performed with the {\textsc Wannier90}\cite{wannier90}  
program on a $N_w = 4\times4\times4$
electron-momentum mesh, by including the LUMO, LUMO+1, and LUMO+2
 states. Both the long-range hoppings and inclusion of the first three LUMO's are needed to get 
localized orbitals and Wannierized bands in a very good agreement with
those computed in the PW basis set
in a window of $\pm$ 0.3 eV around the Fermi level. Indeed, the
maximum discrepancy
between the ab-initio bands and the Wannierized ones is only 5 meV for
the LUMO and LUMO+1 states, while it is larger (0.05 eV at most) for
the LUMO+2 band, that is however higher in energy.

The spreads $\Omega_n = \langle \tilde{w}_{n\vc{0}} | r^2 |
\tilde{w}_{n\vc{0}} \rangle - 
| \langle  \tilde{w}_{n\vc{0}} | \vc{r} | \tilde{w}_{n\vc{0}} \rangle
|^2 $ of the $n$-th MLWF $\tilde{w}_{n\vc{R}}(\vc{r})$ are 
$11.8$\AA, $13.2$\AA, and $21.1$\AA  
for $n =$ 1, 2, and 3, respectively.
The $n=3$ MLWF orbital is more spread out, and leads to a LUMO+2
molecular orbital more sensitive to the crystal environment, as
highlighted by Fig.~\ref{wannier_orbitals}.

\section{Local and non-local electron-phonon couplings
  in K$_3$picene}
\label{local}

The total electron-phonon coupling is 
$\lambda =  \frac{1}{N_\vc{q}} \sum_{\vc{q}\nu}  \lambda_{\vc{q}\nu}$, 
where $\nu$ is the phonon mode
and $\vc{q}$ is its momentum. The phonon resolved coupling reads:
\begin{eqnarray}
\lambda_{\textbf{q} \nu} & = &  \frac{2}{\omega^2_{\textbf{q} \nu}
  N(0)} \frac{1}{N_\vc{k}} \sum_\vc{k} \sum_{n,m} |g^\nu_{\textbf{k}n,\textbf{k}+\textbf{q}m}|^2 
\nonumber \\
   & \times &  (f_{\textbf{k}n}-f_{\textbf{k}+\textbf{q},m}) ~ \delta(\epsilon_{\textbf{k}+\textbf{q},m}-\epsilon_{\textbf{k}n} - \omega_{\textbf{q}\nu}),
\label{elphon}
\end{eqnarray}
that couples the occupied state $|\textbf{k}, n\rangle$ (the ket
refers to the periodic part of the Bloch function) of momentum
$\textbf{k}$ and band $n$ with the empty state $|\textbf{k}+\textbf{q},
m\rangle$ separated by the phonon energy  $\omega_{\textbf{q}\nu}$.
$N(0)$ is the electron DOS per spin per cell at the Fermi level.
The electron-phonon matrix elements are $g^\nu_{\textbf{k}n,\textbf{k}+\textbf{q}m}=
\sum_s \textbf{e}^s_{\textbf{q}\nu} \cdot
\textbf{d}^s_{mn}(\textbf{k}+\textbf{q},\textbf{k})/\sqrt{2 M_s
  \omega_{\textbf{q}\nu}}$, 
where 
$ \textbf{d}^s_{mn}(\textbf{k}+\textbf{q},\textbf{k}) =  \langle
\textbf{k}+\textbf{q},m | \delta v_\textrm{SCF}/ \delta
u_{\textbf{q}s}|\textbf{k},n\rangle$,
with $ \delta v_\textrm{SCF}/ \delta u_{\textbf{q}s}$ the periodic
part of the DFT screened deformation potential.
In Eq.~\ref{elphon}, $f_{\textbf{k}n}$ are Fermi functions depending
on the temperature $T$, and the expression for
$\lambda_{\textbf{q}\nu}$ has to be evaluated 
by a $T \rightarrow 0$  extrapolation.
In the  ``adiabatic'' limit, namely for $\omega_{ph} \ll
\Delta\epsilon$, where $\Delta \epsilon$ is the bandwidth and
$\omega_{ph}$ is the characteristic phonon frequency,
the expression for $\lambda_{\textrm{q}\nu}$ in Eq.~\ref{elphon} reduces to the one proposed by Allen\cite{allen}, and generally used in
previous electron-phonon estimates:
\begin{eqnarray}
\lambda^{AD}_{\textbf{q} \nu} & = &  \frac{2}{\omega_{\textbf{q} \nu}
  N(0)} 
\frac{1}{N_\vc{k}} \sum_\vc{k}
\sum_{n,m} |g^\nu_{\textbf{k}n,\textbf{k}+\textbf{q}m}|^2   
\nonumber \\ 
  & \times & \delta(\epsilon_{\textbf{k},n})  \delta(\epsilon_{\textbf{k}+\textbf{q},m}).
\label{elphon_allen}
\end{eqnarray}
We are going to dub $\lambda^{AD}$ in the Equation above as ``adiabatic'', while $\lambda$ 
in Eq.~\ref{elphon} as ``non-adiabatic''. 

By exploiting the definition of Wannier functions in Eq.~\ref{wannier_def},  the deformation potential matrix elements
$\textbf{d}^s_{mn}(\textbf{k}+\textbf{q},\textbf{k})$
can be written in terms of the molecular MWLF basis as
\begin{eqnarray}
& \textbf{d}^s_{mn}(\textbf{k}+\textbf{q},\textbf{k}) =\sum_\vc{R}
\sum_{m^\prime n^\prime} e^{i \vc{k} \cdot \vc{R}}
\nonumber \\
& U_{m m^\prime}(\textbf{k}+\textbf{q}) \textbf{d}^{\textbf{q}s}_{m^\prime
  n^\prime}(\textbf{R}) U^*_{n n^\prime}(\textbf{k}), 
\label{def_pot_k}
\end{eqnarray}
where  the deformation potential in the MLWF
  local representation is
\begin{equation}
\textbf{d}^{\textbf{q}s}_{m n}(\textbf{R}) =
\langle w_{m\vc{0}} |\frac{\delta v_\textrm{SCF}}{\delta \vc{u}_{\vc{q}s}}
  | w_{n\vc{R}} \rangle.
\label{def_pot_loc}
\end{equation}
Eq.~\ref{def_pot_loc} is the analogous of Eq.~\ref{H_wannier}
but for the electron-phonon coupling elements, when only the
localization of the wave function is used. Therefore, the same
analysis carried out in Sec.~\ref{wannier} can be done here, with the
distinction between the ``local'' (with
$\vc{R}=\vc{0}$ and $(m,n)$ orbitals on a single molecule) and
``non-local''
(with $\vc{R} \ne \vc{0}$, or $\vc{R}=\vc{0}$
with $(m,n)$ orbitals centered on two different molecules of the unit
cell) matrix elements. Thus, it is the Wannier function formalism which
allows one to make the bridge from the plane wave representation
to the molecular orbital description of the electron-phonon problem
introduced in Sec.~\ref{definition}, with the distinction between
local and non-local couplings.

\begin{figure*}[t!h]
\includegraphics[width=2\columnwidth]{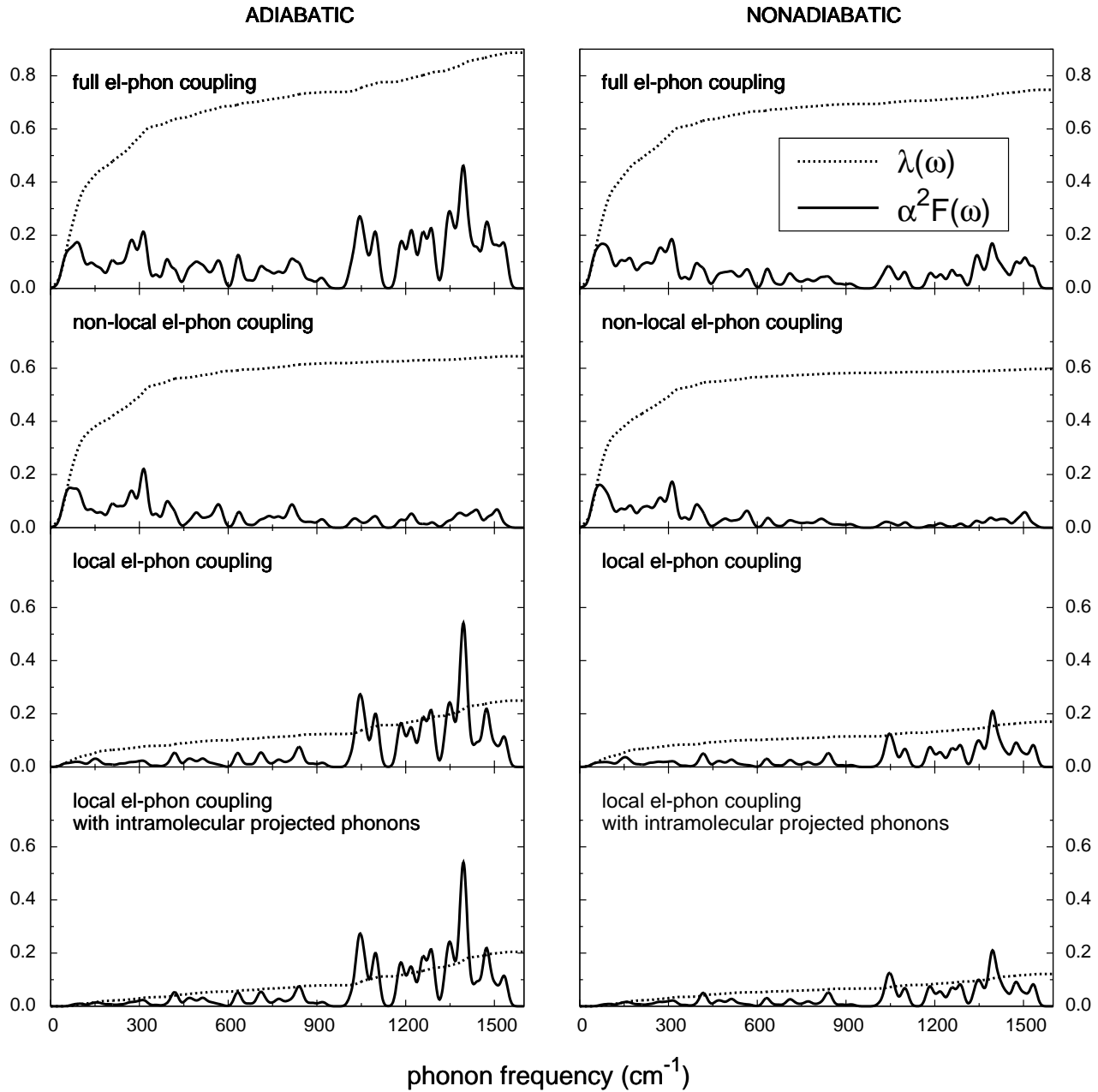}
\caption[]{Eliashberg $\alpha^2 F$ and integrated $\lambda$ for
  various electron-phonon coupling models. All ab-initio elements are
  taken in the first row, the second row is for the non-local $\textbf{d}^{\textbf{q}s}_{m^\prime
  n^\prime}(\textbf{R})$ terms, the third and fourth rows are for the
on-site coupling. In the latter also the phonon eigenmodes are
projected on the intramolecular subspace. To the left we show
quantities computed by the adiabatic approximation
(Eq.~\ref{elphon_allen}), to the right those
evaluated by Eq.~\ref{elphon}.}
\label{intra_and_inter_couplings}
\end{figure*}

As already pointed out in Sec.~\ref{definition}, an analogous but
independent definition of local and non-local contributions can be 
done not only for the electronic states, but also for the phonon modes.
To project
the phonon vibrations we use the same strategy as reported 
in Ref.~\onlinecite{picene_prl_casula}, namely we introduce a 3N$\times$3N
tensor $\mathcal{P}_\textit{S}$, which projects on either the
intramolecular modes or the ensemble of K and intermolecular
modes. $\mathcal{P}_\textit{S}$ acts on the 3D eigenphonons
$\textbf{e}_{\textbf{q}\nu}$, such that one can define the
phonon-projected matrix elements as $g_{\textit{S}} =  \sum_s (\mathcal{P}_\mathit{S} \textbf{e}_{\textbf{q}\nu})^s \cdot
\textbf{d}^s_{nm}(\textbf{k},\textbf{k}+\textbf{q})/\sqrt{2 M_s
  \omega_{\textbf{q}\nu}}$. The resulting phonon-projected $\lambda$
is then
\begin{eqnarray}
\lambda^{\textit{S},\textit{S}^\prime}_{\textbf{q} \nu} & = &
\frac{2}{\omega^2_{\textbf{q} \nu} N(0)} \frac{1}{N_\vc{k} } 
\sum_{\textbf{k},n,m} g_{\textit{S}} ~ g^\star_{\textit{S}^\prime}
\nonumber \\
   & \times &  (f_{\textbf{k}n}-f_{\textbf{k}+\textbf{q},m}) ~ \delta(\epsilon_{\textbf{k}+\textbf{q},m}-\epsilon_{\textbf{k}n} - \omega_{\textbf{q}\nu}).
\label{elphon_proj}
\end{eqnarray}
The total $\lambda$  is $\sum_{\textit{S},\textit{S}^\prime}
\lambda^{\textit{S},\textit{S}^\prime}=\sum_{\textit{S},\textit{S}^\prime}
\frac{1}{N_\vc{q}} \sum_{\vc{q} \nu} \lambda^{\textit{S},\textit{S}^\prime}_{\textbf{q} \nu}$.
The contribution of each subspace $\textit{S}$ is computed as  $\sum_{\textit{S}^\prime} \lambda^{\textit{S},\textit{S}^\prime}$,
where we add both the diagonal term and the usually very small
off-diagonal contributions. 
The results of this
analysis are reported in 
Tab.~\ref{lambda_loc} and
Fig.~\ref{intra_and_inter_couplings}.

\begin{table}[h]
\caption{Adiabatic $\lambda^\textrm{AD}$ and non-adiabatic $\lambda$ computed via
  Eqs.~\ref{elphon_allen} and \ref{elphon}, respectively, for selected
  electron-phonon couplings, corresponding to
  Fig.~\ref{intra_and_inter_couplings}. We report also the phonon
  frequency logarithmic average $\omega _\textrm{log}$ for both the
  adiabatic and non-adiabatic formulations. 
``full el-phon'' means that all terms
  are taken from the ab-initio 
  calculation of the electron-phonon coupling, ``local el-phon'' means that 
that only local terms are retained in $\textbf{d}^{\textbf{q}s}_{m^\prime
  n^\prime}(\textbf{R})$, while ``non-local el-phon'' refers to
the case where only off-site terms are taken in  $\textbf{d}^{\textbf{q}s}_{m^\prime
  n^\prime}(\textbf{R})$. In the ``local el-phon with intra phonons'' not only the deformation potential but
also the phonon eigenmodes are projected on the molecule.
}
\label{lambda_loc}
\begin{ruledtabular}
\begin{tabular}{| l | d | d | r | r |}
\makebox[130pt][c]{model} & \makebox[0pt][c]{$\lambda^\textrm{AD}$} &
\makebox[0pt][c]{$\lambda$} &
\makebox[20pt][c]{$\omega^\textrm{AD}_\textrm{log}$} &
\makebox[20pt][c]{$\omega_\textrm{log}$} \\
\makebox[130pt][c] & \makebox[0pt][c] &
\makebox[0pt][c] &
\makebox[20pt][c]{(meV)} &
\makebox[20pt][c]{(meV)} \\
\hline
\makebox[130pt][c]{full el-phon} & 0.88 & 0.73 & 25 & 18 \\
\makebox[130pt][c]{non-local el-phon} & 0.65 & 0.60  & 17 & 14 \\
\makebox[130pt][c]{local el-phon} & 0.25 & 0.17 & 63 & 41 \\
\makebox[130pt][c]{local el-phon with intra phonons} & 0.20 & 0.12  & 93
& 68 \\
\end{tabular}
\end{ruledtabular}
\end{table}

In Fig.~\ref{intra_and_inter_couplings} we plot the Eliashberg function
$\alpha^2 F(\omega) = \sum_{\textbf{q}\nu} \lambda_{\textbf{q}\nu}
\omega_{\textbf{q} \nu} ~ \delta(\omega - \omega_{\textbf{q} \nu}) /
(2N_q)$, and the integral 
$\lambda(\omega)= 2 \int_0^\omega d\omega^\prime \alpha^2
F(\omega^\prime)/\omega^\prime$, namely the frequency resolved electron-phonon coupling.
The first row is the total $\lambda$ computed by means of
Eqs.~\ref{elphon_allen} and \ref{elphon} by including all
(local and non-local) electron-phonon
contributions. This result has been already reported in
Ref.~\onlinecite{picene_prl_casula}. In the second row, we
plot $\lambda$ where only non-local matrix elements are
taken. This accounts for the $72\%$ ($80\%$) of the total $\lambda$ in
the adiabatic (nonadiabatic) formulation. 
In this case, the main contribution to the
electron-phonon coupling comes from intermolecular
soft phonon modes, with strong
spectral weight at low frequencies ($< 500 \textrm{cm}^{-1}$). 
In the third row of Fig.~\ref{intra_and_inter_couplings}, we plot the
coupling arising only from local matrix elements.
The corresponding $\alpha^2 F(\omega)$ is peaked around frequencies
related to in-plane molecular phonon modes, which give the main
contribution to the local electron-phonon coupling.
In the lower row, not only the deformation potential matrix elements
but also the phonon modes are projected on each molecule. The band
structure ($\epsilon_{\textbf{k}n}$)  and the phonon spectrum
($\omega_{\textbf{q}\nu}$)  are instead unmodified with respect to the
full solid. Therefore,
the latter case is the closest estimate of the
electron-phonon coupling of a single (doped) molecule placed in the crystal
metallic environment.
The projection of the phonon eigenmodes on the molecular
subspace further reduces $\lambda$. We reach therefore one of the main
conclusions of this work. The purely molecular contributions (filtered
in both the wave function and eigenphonons) are such that the resulting $\lambda$
accounts for only $23\%$ ($17\%$) of the full $\lambda$ in the
adiabatic (non-adiabatic) formulation. All the rest comes from
non-local sources. Note also in Tab.~\ref{lambda_loc} that the phonon
frequency logarithmic average $\omega_\textrm{log}$ varies substantially with respect to the
model. The non-local contributions drastically reduce the
$\omega_\textrm{log}$ value, that corresponds to a lower estimate of
$T_c$, which depends linearly on $\omega_\textrm{log}$, according to
McMillan\cite{mcmillan}.

\subsection{Technical details for the electron-phonon calculations}
\label{tech_phon}

In order to evaluate the electron-phonon coupling in the \k3picene,
we first carried out phonon calculations in the density functional
perturbation theory framework (DFPT)\cite{baroni} on a $N_\vc{q}=2\times2\times2$
grid of phonon momenta $\vc{q}$, as illustrated in
Ref.~\onlinecite{picene_prl_casula}. 
The electronic grid used in DFPT has a mesh of $2 \times 2 \times 2$ points
with a Methfessel-Paxton smearing of 0.03 Ry.
For each phonon mode $\nu$ with
momentum $\vc{q}$ we computed both the ``nonadiabatic''
electron-phonon interaction in Eq.~\ref{elphon}, and the ``adiabatic'' one
in Eq.~\ref{elphon_allen}. The $\vc{k}$ summation in both Equations
has been performed by means of the Wannier interpolation
technique\cite{marzari,wannier90,matteo_wannier,giustino_wannier} in the
MLWF basis. The convergence in the $\vc{k}$ summation was reached 
for a $N_\vc{k}=60\times60\times60$ momentum grid, with temperature
and smearing given by $T=150$ K and $\sigma=4.3$ meV, respectively,
with $T=3 \sigma$. 

The matrix elements
$g^\nu_{\textbf{k}n,\textbf{k}+\textbf{q}m}$ have been computed by
selecting the local and non-local contributions in the deformation
potential (through the MLWF via Eq.~\ref{def_pot_loc}) and the phonon modes (through the
phonon projector  $\mathcal{P}_\textit{S}$). 

With the smearing and $\vc{k}$-point grid at convergence, we checked the
accuracy of the extrapolated value of the Fermi level $\epsilon_F$ by
comparing it to the value obtained by an ab-initio calculation with
a large $\vc{k}$-grid ($N_\vc{k}=10\times10\times10$) and a small
Gaussian smearing (0.002 Ry). The uncertainty in $\epsilon_F$ is of the order of 1
meV, giving rise to a change in the density of state $N(0)$ of
1$\%$ only. Thus, a comparable error occurs in $\lambda$ from the $\epsilon_F$  position.

Thanks to the Wannier interpolation the electronic
$\vc{k}$-summation in the electron-phonon coupling $\lambda_\vc{q}$  is converged  
for each $\vc{q}$. 
The main residual error in the total $\lambda$
comes from the coarse $N_\vc{q}=2\times2\times2$ momentum grid used in
the $\vc{q}$-summation of $\lambda_\vc{q}$. An estimate of this error was done in
Ref.~\onlinecite{picene_prl_casula}, by studying the fluctuations of
$\lambda_\vc{q}$ over the $\vc{q}$-point sampling. We found 
that the uncertainty on $\lambda$ goes from about $20 \%$ for the
adiabatic values to less than $15 \%$ for the non-adiabatic estimates,
while the error on the relative contributions (i.e. ratio of local
$\lambda$ over non-local $\lambda$) is even smaller (less
than  $10 \%$).

\section{Screening of local electron-phonon couplings}
\label{mol_screening}

The result presented in the previous Section is completely unexpected,
if one follows the common wisdom that molecular crystals can be
reliably described by molecular derived quantities\cite{Okazaki2010,Ciuchi2012}. For instance,  
molecular electron-phonon calculations have been used to compute $\lambda$ in molecular
crystals\cite{Devos1998,Kato2002}. This has been the case of some previously published works 
on the doped
C$_{60}$\cite{Schluter1992,Devos1997,Han2003,Janssen2010,Faber2011},
and the newly discovered ``aromatic''
superconductors\cite{Devos1998,Kato2002,Kato2002b,Kato2004,Kato2011}. 
Here, we show that at least in the \k3picene, molecular only calculations
are not reliable to predict the crystal total $\lambda$. Presumably, this applies
also to the whole series of new aromatic superconductors,
where the physics should be similar\cite{KubozonoCoronene}.

To explain why the intramolecular electron-phonon coupling is so weak
in the crystal, we carried out electronic structure and phonon calculations for
the isolated neutral molecule in the same  {\textsc quantum-espresso}\cite{qe}
PW framework.
The resulting electron-phonon coupling values are in
good agreement with previous molecular calculations by T. Kato\cite{Kato2011}.

The isolated molecule calculations allowed us to compute also the \emph{molecular}
deformation potential in the MO representation:
\begin{equation}
\textbf{d}^{s ~MOL}_{m n} =
\langle w^{MOL}_m |\frac{\delta V^{MOL}_\textrm{SCF}}{\delta \vc{u}_{s}}
  | w^{MOL}_n \rangle,
\label{def_pot_loc_mol}
\end{equation}
where now $\textbf{d}^{s ~MOL}_{m n}$ is $\vc{q}$-independent.
By replacing $\textbf{d}^{\vc{q}s}_{m n}(\textbf{0})$ with
$\textbf{d}^{s ~MOL}_{m n}$ in
Eq.~\ref{def_pot_k}, and by taking only the local contributions
(namely $\vc{R}=\vc{0}$ and $(m,n)$ running on the same molecule), one can compare
directly the difference between $\frac{\delta
  V^{MOL}_\textrm{SCF}}{\delta \vc{u}_{s}}$ and $\frac{\delta
  v_\textrm{SCF}}{\delta \vc{u}_{\vc{q}s}}$ on the resulting
$\lambda$. Indeed, we have already shown that the MOs $w^{MOL}_m$ are very close
to the molecular MLWF $w_{m\vc{R}}$ (see Fig.~\ref{wannier_orbitals}),
so that a difference in $\lambda$ can come only from the deformation
potential operator. Moreover, we noticed that the $\vc{q}$-dependence
of $\textbf{d}^{\vc{q}s}_{m n}(\textbf{0})$ is very weak, and so a
direct comparison can be made at each crystal phonon momentum $\vc{q}$.
The molecular $\frac{\delta
  V^{MOL}_\textrm{SCF}}{\delta \vc{u}_s}$ is the ``bare'' one,
while $\frac{\delta v_\textrm{SCF}}{\delta \vc{u}_{\vc{q}s}}$ is screened
by the crystal environment and by the partially occupied metallic
bands of LUMO+1 character.

\begin{figure}[t!h]
\includegraphics[width=\columnwidth]{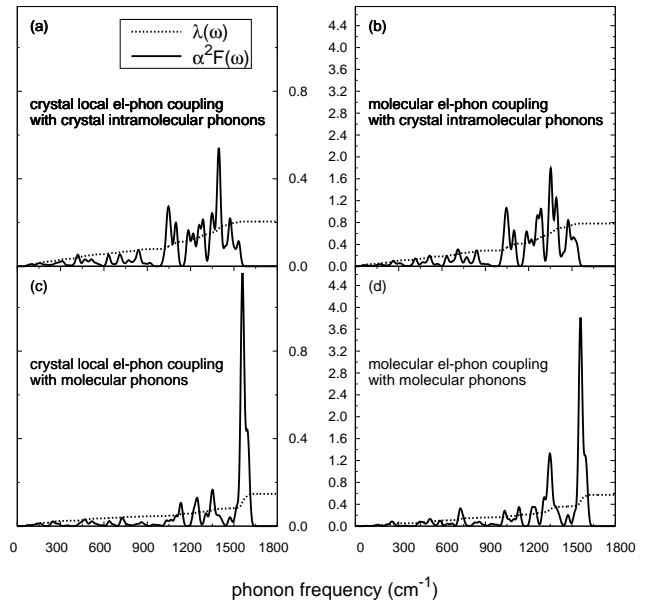}
\caption[]{Eliashberg function $\alpha^2F$ and the adiabatic
  $\lambda^\textrm{AD}$ computed via Eq.~\ref{elphon_allen} for
  both crystal local $\vc{d}$ (panels (a) and (c)) and
  purely molecular $\vc{d}^{MOL}$ (panels (b) and (d)). In the formula,
  we used either the intramolecular projected phonons
  $\vc{e}_{\vc{q}\nu}$ (panels (a) and (b)) or the purely molecular
  dynamical matrix (panels (c) and (d)). Note that the y axis scale of
  the left panels is 4 times wider than the scale on the right panels,
  to show that the magnitude of the local $\vc{d}$ (averaged over the phonon
  momenta $\vc{q}$ and the phonon modes $\nu$) is about 4 times smaller
  than the magnitude of $\vc{d}^{MOL}$ (averaged over the molecular phonon modes
  $\nu$).}
\label{local_screening}
\end{figure}

For here on, our estimates of $\lambda$
are only based on the adiabatic
approximation in Eq.~\ref{elphon_allen}. Although we have seen that it
is less accurate than the nonadiabatic formula, we are going to use it
because from the adiabatic formulation
it is easier to make
the connection to the molecular approximation formula\cite{lannoo} for $\lambda$,
and make the comparison with previous works
(which mainly used the adiabatic approximation).
The results are plotted in
Fig.~\ref{local_screening} and reported in Tab.~\ref{lambda_screening_table}.
By comparing panels (a) and (b), it turns
out that the total coupling $\lambda$ with the screened
deformation potential is about 4 times weaker, which implies that on average the
electron-phonon matrix elements
$g^\nu_{\textbf{k}n,\textbf{k}+\textbf{q}m}$ are twice smaller than
the ``bare'' ones of the isolated neutral molecule.

Therefore, we reach our second main conclusion of this
work. Describing correctly the 
effect of the metallic screening provided by the crystal environment
to the deformation potential is 
critical to get the right estimate of the electron-phonon coupling.

Now, let us analyze in details the effect of the metallic crystal environment on the
dynamical matrix, and so on the phonons. In
Eq.~\ref{elphon_allen}, we replace
the phonon eigenvalues 
$\omega_{\textbf{q}\nu}$ and eigenvectors
$\textbf{e}_{\textbf{q}\nu}$, with the corresponding molecular
$\omega^{MOL}_\nu$ and $\textbf{e}^{MOL}_\nu$, computed for the
isolated undoped picene molecule. The results are
reported in Figs.~\ref{local_screening}(c) and
\ref{local_screening}(d). If compared to panels (a) and (b), 
there is a global frequency softening of 50 cm$^{-1}$ for the in-plane
phonons in the crystal induced by the doping. The second effect is a remodulation of the
frequency dependence of the electron-phonon coupling strength.
In the $\alpha^2 F(\omega)$ obtained with molecular phonons, the coupling is
mostly peaked around 1600 cm$^{-1}$, while it is much more broadly
distributed in the crystal phonons.

From this analysis we can conclude that upon doping the metallic environment
provided by the crystal strongly affects both the deformation
potential and the dynamical matrix. The metallic
screening reduces the electron-phonon coupling strength, while 
it softens the phonon modes and makes their coupling to the charge broader in
the phonon frequency.

The $\alpha^2 F(\omega)$ plotted in Fig.~\ref{local_screening}(d) for
molecular phonons and molecular $\frac{\delta V^{MOL}_\textrm{SCF}}{\delta \vc{u}_{s,0}}$
closely resembles the one published in Ref.~\onlinecite{Subedi}, where
the deformation potential and the dynamical matrix have been computed
for the undoped insulating picene crystal in the rigid doping approximation.
Thus, the effect of
the metallic screening from partially filled bands has been neglected
in both the deformation 
potential and the dynamical matrix. The value of 
$\omega^\textrm{AD}_\textrm{log}$ corresponding to the Eliashberg
function of Fig.~\ref{local_screening}(d) is 125 meV, very close to
the value reported in  Ref.~\onlinecite{Subedi} (126 meV). 
This is a further indication that erroneous results can be obtained
for doped picene if
the metallic screening is not included in the calculations.

\begin{table}[h]
\caption{$\lambda^{AD}$ corresponding to the integrated  $\alpha^2F$
  functions plotted in Fig.~\ref{local_screening}. Row order corresponds to the label sequence of the Figure.
The phonon frequency logarithmic average $\omega^\textrm{AD}_\textrm{log}$ is also reported.
}
\label{lambda_screening_table}
\begin{ruledtabular}
\begin{tabular}{| l | c | r |}
\makebox[195pt][l]{model} & \makebox[0pt][c]{$\lambda^\textrm{AD}$}  &
\makebox[20pt][c]{$\omega^\textrm{AD}_\textrm{log}$}\\
\makebox[195pt][l] & \makebox[0pt][c]  &
\makebox[20pt][c]{(meV)}\\
\hline
\makebox[195pt][l]{crystal local el-phon with crystal intra phonons} &
0.20 & 93 \\
\makebox[195pt][l]{molecular el-phon with crystal intra phonons} &
0.78 & 96 \\
\makebox[195pt][l]{crystal local el-phon with molecular phonons} &
0.15 & 110 \\
\makebox[195pt][l]{molecular el-phon with molecular phonons} & 0.57 & 125 \\
\end{tabular}
\end{ruledtabular}
\end{table}

\subsection{Technical details for the molecular electron-phonon
  calculations}
\label{tech_mol}

The molecular DFT calculations have been carried out with the PW
basis set in the same supercell as the one of the \k3picene, where only one of the two
molecules per crystal unit cell has been taken. 
We checked that the \k3picene supercell is large enough to get the
same molecular levels as the ones of a much larger
supercell, and thus the boundary effects are negligible.
We left the atomic positions of the molecule unchanged from the
crystal, in such a way that the deformation potential calculated for
the molecule could directly replace the one for the crystal in 
$g^\nu_{\textbf{k}n,\textbf{k}+\textbf{q}m}$ of
Eqs.~\ref{elphon} and \ref{elphon_allen} without
any particular rotation in the coordinate space.
The DFT calculation of the molecule was performed in its neutral
state, at the $\Gamma$ point. The electron-phonon calculations were
performed at $\vc{q}=(\frac{1}{2},\frac{1}{2},\frac{1}{2})$ (in crystal fractional
coordinates), to avoid the effective charge contributions to the
deformation potential, that results in the Fr\"ohlich
Hamiltonian\cite{Frohlich1954} and diverges for zone-center
optical phonons.

\section{Impact of dimensionality on non-local electron-phonon couplings}
\label{dimg}

In this section we want to go beyond the distinction between
intramolecular local
and  intermolecular non-local couplings, and analyze what are the most important
electron-phonon interactions among the non-local contributions. We
keep a ``direct space'' approach in labeling the various terms, by
exploiting the  local picture provided by the molecular MLWFs. In
other words, we aim at finding the minimal
electron-phonon lattice model 
(where each lattice site represents the
center of a picene molecule),
which gives the closest possible description
to the full ``ab-initio'' Hamiltonian.

In practice, we select a subset of possible elements in the
deformation potential matrix expressed in the molecular MLWF basis
(Eq.~\ref{def_pot_loc}). According to the set of Bravais vectors $\vc{R}$ and
wannier function indexes $(m,n)$, it is possible to restrict the
coupling to be local (0D), unidimensional (1D) with molecular chains
oriented along the $b$ crystallographic axis, bidimensional (2D) with
molecular layers spanning planes containing the $a$ and $b$
crystallographic axes, or the full ``ab-initio'' model without
constraints. It is also possible to select the neighboring molecules based on their
distance, therefore distinguishing between nearest neighbors (NN) and
next-nearest neighbors (NNN) on a given direction. In
Sec.~\ref{wannier}, we introduced the 1D and 2D models for the
hoppings. The same models apply also for the deformation potential
matrix elements.

\begin{figure}[t!h]
\includegraphics[width=\columnwidth]{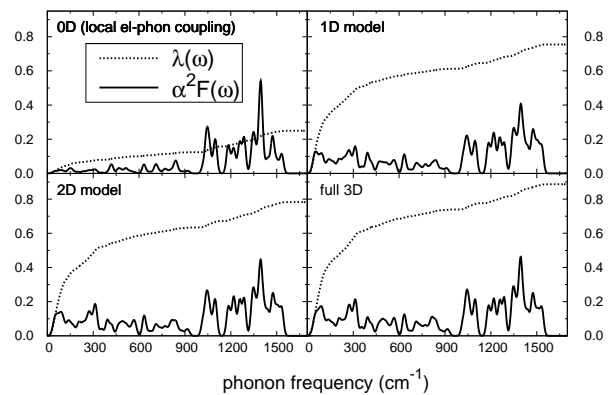}
\caption[]{Eliashberg function $\alpha^2F$ computed with the adiabatic formulation
  (Eq.~\ref{elphon_allen})  for various electron-phonon coupling
  models, based on a selection of the deformation potential $\vc{d}(\vc{R})$ terms.
The 1D model
gives $85\%$ of the total
  $\lambda$, by keeping almost all of the dominant low-frequency contributions,
  related to the coupling with the intermolecular phonons.
}
\label{dimg_fig}
\end{figure}

The results are reported in Fig.~\ref{dimg_fig} and Tab.~\ref{lambda_dim_table}.
The 1D model gives 85$\%$ of the total $\lambda$. The 2D model,
where also the 2D NN and 2D NNN contributions are added, yields 90$\%$ of the total
electron-phonon coupling. From the $\alpha^2 F(\omega)$ in
the upper-rightmost panel of Fig.~\ref{dimg_fig}, it is apparent that in the 1D model the
strongest coupling originates from out-of-plane vibrations and
intermolecular phonons, as a large contribution comes from frequencies
below 300 cm$^{-1}$.

This is the third important result of this work. One can model the
system by \emph{few} non-local electron-phonon couplings
(Eq.~\ref{non_local_coupling}) added to the 
local Holstein-like terms (Eq.~\ref{local_coupling}). As the local terms are weak (see
Sec.~\ref{mol_screening}), those few non-local couplings are
responsible for more than 60$\%$ of the total $\lambda$. This opens
the way toward an efficient and reliable lattice modelization of the
system, where more sophisticated many-body techniques can be used to
deal with the electronic correlation and electron-phonon coupling
together.

It is striking that the main contribution to the coupling (more than 50$\%$)
comes from the phonon modulated hoppings in the $b$ crystallographic direction.
It means that the electron-phonon coupling is
strongly anisotropic along molecular chains. Based on the molecular
arrangement of the crystal and on the band structure, one would have instead expected a planar
anisotropy, as the system is layered, with planes oriented in the
$a$-$b$ directions. 
On the contrary, there is no
clear distinction in magnitude between the in-plane and the
out-of-plane matrix elements, except for the hierarchy between the
strong 1D components and the rest. 
From this point of view, the
material behaves more like an array of chains rather than an array of planes.
The 1D anisotropy is a consequence of the asymmetry of the picene
armchair structure, as already shown in Sec.~\ref{wannier}.

In order to generalize this argument to the experimental situations
and other superconducting aromatic crystals, caution must be taken in
view of the importance of the arrangement of the molecules, and of
disorder. Even though the connection between molecular shape
and electron-phonon anisotropy is intrinsic, and thus
disorder-independent, disorder
in experiments can affect the local geometry and change the total
electron-phonon coupling strength.  

\begin{table}[h]
\caption{$\lambda$ computed via Eq.~\ref{elphon_allen} for various
  electron-phonon coupling models, corresponding to
  Fig.~\ref{dimg_fig}. $\omega^\textrm{AD}_\textrm{log}$ is the phonon
  frequency logarithmic average.
``0D'' means that only local
  molecular couplings are retained in the deformation potential matrix
  elements, ``1D''  refers to the one-dimensional model of chains along the $b$
  crystallographic axis, ``2D'' is the model for the molecular
  herringbone layer spanning the
  $a$ and $b$ axes. 
}
\label{lambda_dim_table}
\begin{ruledtabular}
\begin{tabular}{| l | d | c|}
\makebox[100pt][c]{model} & \makebox[0pt][c]{$\lambda^\textrm{AD}$}  &
\makebox[45pt][c]{$\omega^\textrm{AD}_\textrm{log}$ (meV)}\\
\hline
\makebox[100pt][c]{0D} & 0.25 & 63\\
\makebox[100pt][c]{1D} & 0.74 & 26\\
\makebox[100pt][c]{2D} & 0.77 & 27 \\
\makebox[100pt][c]{full 3D} &  0.88  & 25 \\
\end{tabular}
\end{ruledtabular}
\end{table}

\section{Conclusions}
\label{conclusions}

In this paper we went beyond what we have done in
Ref.~\onlinecite{picene_prl_casula}, where we carried out a detailed analysis of
the electron-phonon coupling based on the phonon projection to intramolecular and
intermolecular eigenmodes. Here, thanks to the maximally localized Wannier functions
formalism, we took into account also the 
electronic projection of the deformation potential into local
(Holstein-like) and non-local couplings, defined based on a molecular
orbital representation. 
We found that the purely molecular contribution (projected in both the
deformation potential and phonon eigenmodes) is very weak. It accounts for only $20\%$ of the total $\lambda$
(in the adiabatic coupling formulation of Eq.~\ref{elphon_allen}).
It is therefore impossible to predict the total coupling of \k3picene with isolated molecular
calculations only. 
We understood
this as an effect of the metallic screening which mainly reduces the deformation
potential matrix elements.
 This turns out from a direct comparison of
the projected local coupling in the crystal with the full coupling in
the isolated molecule, carried out within the same theoretical
framework.

Moreover, we demonstrated that while $80\%$ of 
the electron-phonon coupling in \k3picene 
is non-local, more than $60\%$ of $\lambda$ comes just from \emph{two}
terms, i.e. the 1D nearest neighbors (NN) and 
the 1D next-nearest neighbors (NNN) phonon-modulated hoppings, which form
ladder chains along the $b$ crystallographic axis of the compound.

The strong spatial $1D$ anisotropy of the electron-phonon coupling is another
interesting outcome of the present work. The intermolecular modes couple more
strongly with the electrons along chains of molecules, arranged in an ordered
array of ladders. We related this to the picene molecular edge asymmetry
in its armchair structure. 

Finally, we showed that the model comprising of local, 1D NN and 1D NNN contributions yields $85\%$ of the total
electron-phonon coupling of the crystal. This \emph{ab initio}
modelization 
opens the way to reliable
and quantitative many-body calculations on the lattice, in order to study the
interplay between strong electronic correlation (typical of a
molecular crystal with flat bands\cite{Giovannetti,Kim,japanese_wannier}), and
electron-phonon coupling in \k3picene.

\begin{acknowledgments}
We acknowledge M. Fabrizio, T. Kato, Y. Kubozono, G. Profeta, and S. Taioli for useful discussions. 
M. Casula thanks GENCI for the HPC resources obtained under the
Grant 2012096493.
\end{acknowledgments}


\begin{thebibliography}{00}
\bibitem{footnote1} In the text we prefer to give the definitions based on a
  finite summation over the BZ rather than the continuous limit of the
  integral, as in the practical implementation of the formula the finite grid is
  always used. This allows us to specify also the different
  $\mathrm{k}$ grids used for various quantities. In general, one
  can recover the continuous formulation by substituting $\frac{1}{N_w}
  \sum_\mathrm{k}$ with $\frac{V}{(2\pi)^3} \int_{\mathrm{BZ}} \!\!
  \dd\vc{k}$, whenever it appears.
\end{thebibliography}
\end{document}